\documentclass[prb,aps,showpacs,twocolumn]{revtex4}
\usepackage{bm}
\catcode`\"=\active \let"=\" \let\3=\ss
\renewcommand{\d}{\displaystyle}
\newcommand{\Kr}[1]{\left( #1\right)}
\newcommand{\Kg}[1]{\left\{ #1\right\}}
\newcommand{\Ke}[1]{\left[ #1\right]}

\newcommand{\bra}[1]{\raisebox{-0.3ex}{\mbox{\large\tt <}}#1
 \hspace{0.5ex}\rule[-0.5ex]{0.1ex}{0.9em}\hspace{0.5ex}}
\newcommand{\ket}[1]{\hspace{0.5ex}\rule[-0.5ex]{0.1ex}{0.9em}
 \hspace{0.5ex}#1\raisebox{-0.3ex}{\mbox{\large\tt >}}}
\newcommand{\braket}[2]{\raisebox{-0.3ex}{\mbox{\large\tt <}}#1
 \hspace{0.5ex}\rule[-0.5ex]{0.1ex}{0.9em}\hspace{0.5ex}
 #2\raisebox{-0.3ex}{\mbox{\large\tt >}}}
\newcommand{\bms}[1]{\mbox{\scriptsize\boldmath$#1$}}
\begin{document}
\bibliographystyle{revtex}
\title{Spin conversion rates due to dipolar
interactions in mono-isotopic quantum dots
at vanishing spin-orbit coupling}
\author{Wolfgang H\"ausler$^{1,2}$\footnote{Corresponding 
Author} and Peter H\"anggi$^1$}
\address{$^1$Institut f"ur Physik,
Universit"at Augsburg,
Universit"atsstr. 1,
86135 Augsburg, Germany \\
$^2$I.~Institut f"ur Theoretische Physik,
Universit"at Hamburg,
Jungiusstr. 9,
20355 Hamburg, Germany
}
\begin{abstract}
Dipolar interaction between the magnetic moments of electrons is
studied as a source for electron spin decay in quantum dots or
arrays of quantum dots. This magnetic interaction will govern
spin decay, after other sources, such as the coupling to nuclear
spins or spin orbit coupling, have been eliminated by a suitable
sample design. Electron-electron (Coulomb) interactions,
important for magnetic properties, are included. Decomposing
the dipolar operator according to the symmetric group of
electron permutations allows one to deduce vanishing decay
channels as a function of electron number and spatial symmetries
of the quantum dot(s). Moreover, we incorporate the possibility
of rapid phonon induced spin conserving transitions which
crucially affect the temperature dependence of spin decay rates.
An interesting result is that a sharp increase of the spin decay
rate occurs already at relatively low temperatures.
\end{abstract}
\maketitle

\section{Introduction}
One of the proposals to realize qubits for quantum computing
\cite{awschalom02} uses the electron's spin in semiconducting
solid state nanostructures.\cite{LossDiVincenzo,elzerman05} This
approach could benefit from traditional electronics device
experience and allows for straightforward scalability. Recent
efforts have succeeded to demonstrate controlled preparation and
detection of single \cite{elzerman04} and of pairs
\cite{hanson04} of electron spins in quantum dots. Aiming for
long running coherent computations, however, solid state based
devices somewhat suffer from relatively short spin decoherence
and relaxation times as a drawback, for example, compared to
nuclear spin based qubits.\cite{kane98} Exponential decay of the
upper Zeeman level population has been observed in GaAs quantum
dots \cite{elzerman04} over times not exceeding 10$^{-3}\:$sec.
Therefore, it is important to know (and, if possible, to
control) any kind of mechanism causing spin relaxation in
solids, and particularly in semiconductors. Previous
theoretical work has valued several contributions. They can be
subdivided into two classes regarding the magnetic mechanism to
mix spin states as a source for spin decay: (i) spin-orbit
coupling,\cite{khaetskii00,khaetskii01,woods04,golovach04} also,
recently, in its interplay with the electron-electron
interaction,\cite{kavokin,badescu} or (ii) coupling to nuclear
spins. The latter can act through the spin-flip Overhauser
effect by hyperfine
interactions.\cite{erlingsson01,erlingsson02,khaetskii02,abalmassov04}
In principle, both magnetic sources for spin mixing can be
eliminated by a proper device design and by the choice of the
semiconductor material. Non-vanishing spin-orbit splitting can
have several causes in semiconductors which we briefly address:
$p$-type bands, i.e.\ usually valence bands, may split by
spin-orbit effects, arising near the nuclei. Secondly does the
lack of spatial inversion symmetry produce spin splitting even
of $s$-type bands, either by the Dresselhaus mechanism arising
in the absence of crystallographic centrosymmetry as in
Zincblende or Wurtzite structures. The latter particularly
refers to all III-V semiconductors with GaAs being the most
striking example. Also devices lacking structural inversion
symmetry, e.g.\ near surfaces or in asymmetric quantum wells,
producing internal electric fields show spin splitting due to
the Rashba mechanism. Fortunately, this latter spin-orbit
source may be suppressed by fine tuning suitable gate
voltages.\cite{gatevary} The goal to avoid spin-orbit coupling
effects therefore suggests using spins of conduction band
electrons in Si or in Ge in carefully symmetrically prepared
structures. Also the attempt to avoid coupling to nuclear
spins favors the use of Si or Ge: their natural isotopic mixture
contains nuclear spin $I=0$ to more than 95\% (Si) or more than
92\% (Ge), respectively.\cite{frauenfelder}

With this work we consider the effect of dipolar interactions
between the magnetic moments of electron spins which, for
fundamental reasons, {\em cannot\/} be removed by design. While
considerably weaker \cite{abrahams} than the above quoted
mechanisms this interaction unavoidably causes spin relaxation
and, in the absence of other magnetic interactions, combined
with the never vanishing electron-phonon
coupling,\cite{sarmacomment} will set the ultimate limit for
long time quantum computations using electron
spins,\cite{LossDiVincenzo} even in optimally designed
structures. We study transitions between energy levels
differing in their {\em total\/} spins \cite{entanglement} and
disregard here effects associated with transitions between
Zeeman levels (which conserve the symmetry of many body electron
levels, see below) at finite magnetizations when an external
magnetic field is applied. After introducing the model in
Sect.~\ref{model} we reveal circumstances of particular spin
stability w.r.t.\ dipolar interactions
(Sect.~\ref{dipolarenergy}), also regarding excited (many
electron) states in Sect.~\ref{stability}, depending on the
electron number and on the symmetry of the single or the
ensemble of quantum dots.

We explicitly include Coulomb interactions \cite{kavokin,badescu}
due to their importance for magnetic properties. For example,
they can cause total ground state spins greater
\cite{reimann,mak,epl00,mikhailov} than $S=0$ or $S=1/2$, as
expected for even or odd numbers $N$ of non-interacting
electrons. Here, we take the electrons to be confined inside
one quantum dot or in different quantum dots. As a
complementary approach to the non-interacting or weakly
interacting regime we focus on strong Coulomb interactions at
low electron densities, where pocket states,
\cite{fkp,zpb,annalen} offer a reliable description of many body
states localized by Coulomb repulsion, even in single quantum
dots.\cite{mak} Pocket states are briefly reviewed in
Sect.~\ref{pocketstates} for the present purpose to determine
matrix elements of the dipolar interaction in
Sect.~\ref{stability}.

Pursuing most of the foregoing theoretical work on electron spin
decay in quantum dots we consider in Sect.~\ref{electronphonon}
phonons (which themselves cannot change spin states) to provide
the transition energy between discrete dot levels. Contrary to
extended bulk situations, this transition energy is much bigger
than mere magnetic energies which is is one reason for the
relative stability of quantum dot compared to bulk electron
states, in accordance with experimental fact.\cite{stabledots}
Spin changing transitions due to the combined action of dipolar
energy and phonons are discussed in Sect.~\ref{transrate}.
Generalizing previous results we account for rapid {\em spin
conserving\/} excitations of the electron system induced by
phonons that occur already at relatively low temperatures; these
transitions turn out to govern predominantly the temperature
dependence of spin decay times, discussed in
Sect.~\ref{temperaturedependence}. Finally, we summarize and
value our findings in Sect.~\ref{resume}.

\section{Model}\label{model}
Specifically, we consider the $N$-electron system
\begin{equation}\label{h0}
H_0=\sum_{i=1}^N\Kr{\frac{\bm{p}_i^2}{2m^*}+v(\bm{r}_i)}+
\frac{1}{2}\sum_{i\ne j}w(|\bm{r}_i-\bm{r}_j|)
\end{equation}
confined by the potential $v(\bm{r})$ which is supposed to describe
a single quantum dot or more complex situations of many quantum
dots, such as for example $N$ quantum dots each containing
a single electron. To be realistic, particularly regarding
magnetic properties, we include interactions between electrons
$w(r)=\frac{e^2}{\kappa}r^{-1}$, depending on the static
dielectric constant $\kappa$ of the host material; Coulomb
interactions are always considerably stronger than dipolar
energies. Moderate screening, not reducing the interaction
range to values smaller than the electron separation, will not
affect qualitatively our results. In Eq.~(\ref{h0}), $\bm{p}_i$
and $\bm{r}_i$ are $d$-component momentum and position vectors,
depending on the dimensionality $d$ of the quantum dot wave
functions (in heterostructures, $d=2$); $m^*$ is the
band electron mass.

Notice that at strong Coulomb repulsion, which is the focus of
this work, $N$ electrons Wigner crystallized \cite{jauregui,mak}
in a single quantum dot become in their theoretical treatment at
low energies very similar to the case of $N$ electrons localized
in separate quantum dots. The essential physics of both situations
is captured by an antiferromagnetic Heisenberg lattice
model.\cite{fkp,jhjwh} Eigenstates of $H_0$ exhibit well defined
spins $S$ and can be classified according to the eigenvalues of
the z-component $\hat S_z$ and the square $\bm{\hat S}^2$ of the
total spin operator $\bm{\hat S}=\sum_{i=1}^N\bm{\hat S}_i$
yielding eigenvalues $S_z$ and $S(S+1)$, respectively. With
SU(2) symmetry in spin space, Zeeman multiplets $\:-S\le S_z\le
+S\:$ are degenerate. We index eigenstates $\ket{\phi_n}$ and
eigenvalues $E_n$ of $H_0$ by $n$, taken to incorporate the
values of $S$ and $S_z$. Transitions between $\ket{\phi_n}$ and
$\ket{\phi_{n'}}$ may or may not change $S$. The present work
focuses on inelastic transitions that change the {\em total\/}
spin values $S\to S'$ rather than on transitions within a Zeeman
multiplet. As already mentioned, many-electron ground states
$\ket{\phi_{n=0}}$ may exhibit total spin values $S_0>1/2$ as a
result of electron-electron
interactions.\cite{reimann,mak,epl00,mikhailov}

\section{Transition matrix elements}\label{matrixelements}
In order to satisfy the Pauli principle an $N$-fermion state
$\braket{\bm{r}_1,s_1,\ldots,\bm{r}_N,s_N}{\phi_n}$ must belong
to the $A_2\equiv[1^N]$ representation of the symmetric
(permutational) group $S_N$ with respect to permutations $p\in
S_N$ of the particle enumeration,
$\:\{1,\ldots,N\}\to\{p(1),\ldots,p(N)\}\:$, see
Ref.~\onlinecite{hamermesh}. When permuting only spin
coordinates $\{s_1,\ldots,s_N\}\to\{s_{p(1)},\ldots,s_{p(N)}\}$
the state $\ket{\phi_n}$ transforms according to the irreducible
representation (partition) $\Gamma=[N/2+S,N/2-S]$ of $S_N$ for
spin-$\frac{1}{2}$ Fermions \cite{zpb,annalen} at given
$S=\left\{{0\atop 1/2}\right\},\ldots,N/2$ for
${\Kg{\mbox{\footnotesize even}\atop\mbox{\footnotesize
odd}}}\;N$. Correspondingly, when permuting only positions
$\{\bm{r}_1,\ldots,\bm{r}_N\}\to\{\bm{r}_{p(1)},\ldots,\bm{r}_{p(N)}\}$,
$\ket{\phi_n}$ transforms according to
$\bar\Gamma=[2^{N/2-S},1^{2S}]$ (with $\Gamma\times\bar\Gamma$
containing the $A_2$-representation). We notice that total spin
changing transitions require altering the wave function's
symmetry, which necessitates operators acting simultaneously in
position and in spin space (by contrast, transitions within a
Zeeman multiplet leave unaltered the symmetries of wave
functions).

\subsection{Dipolar energy}\label{dipolarenergy}
Here, we investigate the dipolar interaction $H^{\rm D}$ between
electrons. As seen in Eq.~(\ref{hddij}) below, it contains
products of position and spin operators and, indeed, mixes spin
states. However, it is by far too weak to provide the energy

separating quantum dot eigenlevels. Focusing on Si, we
consider in Section~\ref{spinrelaxrates} acoustical deformation
potential phonons \cite{shockley} to supply the necessary
transition energy. Unaided electron-phonon coupling, though,
does not mix spin states and thus leaves spins unaltered.
Eventually, it turns out that dipolar interaction, as a result
of its smallness, ensues considerably smaller transition rates
at low temperatures than, for instance, nuclear spin induced
spin mixing.\cite{erlingsson01}

The operator of the dipolar energy
\begin{equation}\label{hdd}
H^{\rm D}=\frac{1}{2}\sum_{i\ne j}H_{ij}^{\rm D}
\end{equation}
is, as required for identical particles, invariant with respect
to permuting the electron enumeration; however, $H^{\rm D}$ can
be decomposed into parts that are {\em not\/} invariant under
permuting coordinates $\bm{r}_i$ or spins $\bm{\hat S}_i$
separately. Let us first recap the interaction between a
pair of magnetic moments
\begin{equation}\label{hddij}
H_{ij}^{\rm D}=\frac{\gamma^2}{r_{ij}^5}
\Ke{r_{ij}^2\bm{\hat S}_i\cdot\bm{\hat S}_j-
3(\bm{r}_{ij}\cdot\bm{\hat S}_i)(\bm{r}_{ij}\cdot\bm{\hat S}_j)}
\end{equation}
where $\gamma=ge\hbar/2mc$ ($c$ is the velocity of light and the
$g$-factor for dot carriers which even in few electron quantum
dots is found to take basically bulk
values).\cite{rokhinson,hanson04} Its Heisenberg-like first part
is manifestly SU(2)-invariant in spin space and commutes with
$\bm{\hat S}^2$. This part neither changes $S$ nor $S_z$ and
just renormalizes the energies slightly. It therefore can be
ignored in view of the smallness of dipolar energies compared to
the dot level separations. In Eq.~(\ref{hddij}) we abbreviate
$\bm{r}_{ij}:=\bm{r}_i-\bm{r}_j$ and $r_{ij}:=|\bm{r}_{ij}|$.
The second part of $H_{ij}^{\rm D}$ can be decomposed as
\begin{equation}\label{hdij}
\frac{1}{r_{ij}^5}(\bm{r}_{ij}\cdot\bm{\hat S}_i)
(\bm{r}_{ij}\cdot\bm{\hat S}_j)=
\Ke{H_{ij}^{(0)}+H_{ij}^{(1)}+H_{ij}^{(2)}}
\end{equation}
where the three terms
\begin{eqnarray}
H_{ij}^{(0)}&=&\d\frac{|\varrho_{ij}|^2}{4}
\Kr{\hat S_{+i}\hat S_{-j}+\hat S_{-i}\hat S_{+j}}+
\zeta_{ij}^2\hat S_{zi}\hat S_{zj}\label{h0ij}\\
H_{ij}^{(1)}&=&\d\frac{\zeta_{ij}}{2}
\Biggl[\varrho_{-ij}(\hat S_{+i}\hat S_{zj}+\hat S_{zi}\hat S_{+j})\\
&&\mbox{}+\d\varrho_{+ij}(\hat S_{-i}\hat S_{zj}+\hat S_{zi}\hat S_{-j})
\Biggr]\label{h1ij}\nonumber\\
H_{ij}^{(2)}&=&\d\frac{\varrho_{+ij}^2}{4}\hat S_{-i}\hat S_{-j}+
\frac{\varrho_{-ij}^2}{4}\hat S_{+i}\hat S_{+j}\label{h2ij};,
\end{eqnarray}
are responsible to alter symmetries and spins after carrying out
summation over $(i\ne j)$. In Eqs.~(\ref{h0ij}---\ref{h2ij})
they change $S_z$ by $0$, $\pm 1$, and $\pm 2$, respectively.
$\hat S_{\pm}:=\hat S_x\pm{\rm i}\hat S_y$ denote usual rising
or lowering operators in spin space, $\varrho_{\pm
ij}:=(x_{ij}\pm{\rm i}y_{ij})/r_{ij}^{5/2}$ is a complex
(angular momentum generating) coordinate in the plane
perpendicular to the axes of spin quantization, taken as the
$z$-axes, and $\zeta_{ij}:=z_{ij}/r_{ij}^{5/2}$.

The spin changing part Eqs.~(\ref{h0ij}---\ref{h2ij}) of $H^{\rm D}$
can now further be decomposed according to partitions $\Gamma$
of the symmetric group $S_N$,
\begin{eqnarray}\label{decompose}
&&\d\sum_{i\ne j}(H_{ij}^{(0)}+H_{ij}^{(1)}+H_{ij}^{(2)})\\
&&=\d H^{\Gamma=[N]}+H^{\Gamma=[N-1,1]}+H^{\Gamma=[N-2,2]}\;.
\nonumber
\end{eqnarray}
This latter representation is particularly useful to deduce
non-zero transition matrix elements between quantum dot
eigenstates of different {\em total\/} spins. No other
partitions occur since $H_{ij}^{\rm D}$ transforms as a product
of two vector operators in position as well as in spin space,
cf.\ Eq.~(\ref{hddij}), i.e.\ as a tensor of rank two. In
Eq.~(\ref{decompose}) $H^{\Gamma=[N-1,1]}$ changes the total
spin $S$ of dot by $\pm 1$ and $H^{\Gamma=[N-2,2]}$ by $\pm 2$,
where the latter occurs only for $N\ge 4$ while the former
already for $N\ge 3$. $H^{\rm D}$ cannot achieve spin changes
by more than $\pm 2$. For example, we can conclude already at
this stage that the rate for {\em direct\/} transitions of an
excited $S=3$ quantum dot state into the (assumed) $S=0$ singlet
ground state will be of the order ${\cal O}((H^{\rm D})^4)$ and
therefore will be very small. All properly symmetrized operators
$H^{\Gamma=[N-1,1]}$ and $H^{\Gamma=[N-2,2]}$ for $N=3$ and
$N=4$ are listed in Appendix~\ref{n3n4operators}.

Note, that the property of $H_{ij}^{(0-2)}$ to change $S_z$ by
0, $\pm 1$, $\pm 2$, respectively, is unrelated to their
respective capability to change $S$. In the absence of further
symmetries of the quantum dot shape, all three operators
$H_{ij}^{(0-2)}$ contain both, $H^{\Gamma=[N-1,1]}$ and
$H^{\Gamma=[N-2,2]}$. In cases of frozen electron motion in
$z$-direction, as it applies to quantum dots (or arrays of
quantum dots) fabricated on the basis of semiconducting
hetero-structures \cite{leoreview} all of the above
contributions involving $\zeta_{ij}$ vanish. Then $H^{(0)}$
simplifies and $H^{(1)}$ vanishes entirely, so that $S_z$ can
either remain unaltered (through $H^{(0)}$) or change by $\pm 2$
(through $H^{(2)}$).

\subsection{Two electrons}
Let us first focus on two electrons, i.e.\ $N=2$. This is
relevant, for example, for double dots containing one electron
on either side to realize the basic entity of coupled
qubits.\cite{LossDiVincenzo,thorwart02} As already mentioned in
the previous section, non-$A_1$ symmetric partitions
Eq.~(\ref{decompose}) of $H^{\rm D}$ occur only for $N\ge 3$.
Further, $H^{\rm D}$ does not contain the $A_2$ partition for
any $N$. Therefore, spin conversion transitions from a triplet
excited state into the singlet ground state \cite{liebmattis}
will never be mediated by $H^{\rm D}$. In the related physics
context of nuclear spin conversion of H$_2$ molecules
\cite{bonhoeffer} the stability of ortho-hydrogen (even over
weeks) is traced back \cite{silvera} to parity symmetry of both,
the molecule and the magnetic dipolar interactions between the
two protons (of actually close proximity which enhances dipolar
forces) to prevent the transition from the odd-parity ortho
$S=1$ state into the (by 80 Kelvin lower) even-parity para $S=0$
ground state. In this case, spins refer to the protons. In the
context of quantum dots we can generalize this finding:
Irrespective of the shape of the quantum dot confining potential
and of the functional form of the electron-electron interaction
$w(\bm{r}_1-\bm{r}_2)$ the dipolar interaction will not change
(triplet or singlet) spin states as a result of permutational
symmetry and quantum mechanical particle identity of $N=2$
electrons. This statement is not restricted to the lowest
(golden rule) order ${\cal O}((H^{\rm D})^2)$ but even holds
true to any order of $H^{\rm D}$. As one neat corollary we
conclude that two electrons in square shaped quantum dots (in
the absence of other magnetic mechanisms) will stay in their
respective spin states. This supports a corresponding proposal
for quantum computations based on superpositions of states where
the two electrons occupy either of the two equivalent
electrostatic energy minimum positions at diagonally opposite
corners in a square.\cite{jhj}

\subsection{Strong interaction, pocket states}\label{pocketstates}
In case of more than two electrons we focus on strong Coulomb
forces at low carrier densities, i.e.\ at large values of the
electron gas parameter $r_{\rm s}\gg 1$. Then, the kinetic
energy is small and the electron system lowers its energy by
Wigner localizing \cite{jauregui} the charge density near
electrostatically favorable places. Precursors of Wigner
crystallization have been found already at $r_{\rm s}\approx 4$
in two-dimensional quantum dots.\cite{mak} A similar
localization of charge density arises when the external
confining potential separates the electrons, such as in the
case of $N$ quantum dots, each containing a single electron.
In either case, at strong Coulomb interactions, eigenstates
$\ket{\phi_n}$ of $H_0$ Eq.~(\ref{h0}) are well described by
pocket states,\cite{fkp,zpb,annalen} which exploit the electron
localization. They allow to estimate the spin dependent low
energy spectrum to exponential accuracy with increasing $r_{\rm
s}$, or with increasing dot separation.

In the Wigner crystal state, electrons vibrate about
electrostatic energy minimum positions. Linearizing the
(Coulomb and external) forces yields the plasmon spectrum of the
confined $N$-electron system. Energy level separations
$\omega_{\rm pl}^2\sim\omega_0^2+Ar_{\rm s}^{-3}$ can be
estimated from the dynamical matrix with a prefactor $A$
depending on $N$ and on the dot lay-out; $\omega_0$ is the
confining frequency of the quantum dot(s). Due to the electron
spin each plasmon level is $2^N$-fold degenerate. Quantum
corrections (partly) split this degeneracy into sub-levels, with
all exhibiting well defined total spins $S=\Kg{0\atop
1/2},\ldots,N/2$ for $\Kg{\mbox{\rm even}\atop\mbox{\rm
odd}}\;N$ [of $(2S+1)$-fold Zeeman degeneracy, by spin rotation
invariance], according to Sect.~\ref{model}. The ground state
(in more than one spatial dimension) need not be of minimal spin
$S_0=0$ or $S_0=1/2$.\cite{reimann,zpb,mikhailov,epl00} A given
spin $S$ may appear more than once in such a spin split plasmon
level; examples of spectra are discussed in
Refs.~\onlinecite{fkp,zpb,annalen}. The splitting arises due to
permutational electron exchanges by quantum mechanical tunneling
through the electrostatic barrier consisting of the $v$-term
plus the $w$-term in Eq.~(\ref{h0}). In the simplest case there
are $N!$ different, but all energetically precisely equivalent,
possibilities to arrange the localized electrons; this defines
$1\le p\le N!$ pocket states $\ket{p}$. The width of each
pocket state corresponds to plasmonic zero point oscillations
and scales roughly as $\omega_{\rm pl}^{-1/2}$ in
$Nd$-dimensional configuration space ($d$ being the spatial
dimensionality of the quantum dot, often \cite{leoreview} $d=2$
but also $d=1$ is realized, for example in rods of carbon
nanotubes \cite{pablo}). The energy scale $\Delta$ for spin
splittings of plasmon levels through quantum mechanical electron
exchanges is tuned by the magnitude of overlap integrals
$\bra{p'}H_0\ket{p}$ between two different arrangements $p$ and
$p'$. This latter quantity can be estimated semiclassically
\cite{zpb,annalen,jhjwh} to read
$\Delta\sim\bra{p'}H_0\ket{p}\sim\omega_{\rm
pl}\:\exp(-\sqrt{r_{\rm s}})$ so that $\Delta/\omega_{\rm pl}\ll
1$. Numerically obtained quantum dot spectra
\cite{zpb,mak,polygon} indeed nicely follow this behavior. For
example, it exhibits the predicted \cite{epl00} crossover into a
spin polarized $S=3/2$ ground state in a spherical two
dimensional quantum dot containing $N=3$ electrons at
sufficiently low electron density.\cite{mak,mikhailov} As a
result, all eigenstates
\begin{equation}\label{phicp}
\ket{\phi_n}=\frac{1}{N_n}\sum_pc_{np}\ket{p}
\end{equation}
belonging to the plasmon ground multiplet can approximately be
expressed through the set $\{\ket{p}\}$. The (real) coefficients
$c_{np}$, appearing in Eq.~(\ref{phicp}), ensue from the
irreducible representation $\Gamma$ of the permutational group
$S_N$, according to the wave functions symmetry which at the
same time fixes the total spin $S$ of $\ket{\phi_n}$;
$N_n=\sqrt{\sum_{pp'}c_{np}c_{np'}\braket{p'}{p}}$ ensures
normalization, $\braket{\phi_n}{\phi_n}=1$.

\subsection{Dipolar matrix elements}\label{stability}
Pocket states allow to conveniently estimate the matrix elements
$\bra{\phi_{n}}H^{\rm D}\ket{\phi_{n'}}$ of the dipolar energy
since, to leading order, electron positions may be taken as
being well localized, $\delta$-function like, for $\varrho_{ij}$
and $\zeta_{ij}$ in Eqs.~(\ref{h0ij}---\ref{h2ij}) or in the already
symmetrized expressions in Appendix~\ref{n3n4operators}.
This leads to a finite lattice spin problem. Having constructed
symmetrized spin states, the matrix elements of $H^{\rm D}$ for
$N>4$ follow straightforwardly from Eq.~(\ref{decompose}).

We demonstrate our approach for the particularly symmetric cases
of $N=3$ and $N=4$ electrons occupying equilateral electrostatic
equilibrium positions, as in a two-dimensional spherical quantum
dot,\cite{mak,epl00} in triangularly or square shaped quantum
dots,\cite{leoreview} or in equilateral triangular or square
arrangements of single electron quantum dots. We assume frozen
motion in $z$-direction, as in heterostructures, so that terms
involving $\zeta_{ij}$ or $\zeta_{ij}^2$ are irrelevant in
Eqs.~(\ref{h0ij}---\ref{h2ij}). Symmetrized, non-trivial spin
states of minimal $S_z$-components are presented in
Table~\ref{spinstates}.

\begin{widetext}
\newcommand{\up}{\mbox{\footnotesize$\uparrow$}}
\newcommand{\dw}{\mbox{\footnotesize$\downarrow$}}
\begin{center}\begin{table}\begin{tabular}{|c|c|c|c|l|}\hline
$N$&$S$&$S_z$&index&\\ \hline
$3$&$\frac{3}{2}$&$\frac{1}{2}$&A&$\frac{1}{\sqrt{3}}(\ket{\up\up\dw}+
\ket{\up\dw\up}+\ket{\dw\up\up})$\\
$3$&$\frac{1}{2}$&$\frac{1}{2}$&E$_{\rm a}$&
$\frac{1}{\sqrt{3}}(\ket{\up\up\dw}+{\rm e}^{{\rm i}2\pi/3}\ket{\up\dw\up}+
{\rm e}^{-{\rm i}2\pi/3}\ket{\dw\up\up})$\\
$3$&$\frac{1}{2}$&$\frac{1}{2}$&E$_{\rm b}$&
$\frac{1}{\sqrt{3}}(\ket{\up\up\dw}+{\rm e}^{-{\rm i}2\pi/3}
\ket{\up\dw\up}+{\rm e}^{{\rm i}2\pi/3}\ket{\dw\up\up})$\\ \hline

$4$&$2$&$0$&A&$\frac{1}{\sqrt{6}}(\ket{\up\up\dw\dw}+\ket{\dw\up\up\dw}+
\ket{\dw\dw\up\up}+\ket{\up\dw\dw\up}+\ket{\up\dw\up\dw}+
\ket{\dw\up\dw\up})$\\
$4$&$1$&$0$&T$_{\rm x}$&$\frac{1}{2}(\ket{\up\up\dw\dw}-
{\rm i}\ket{\dw\up\up\dw}-\ket{\dw\dw\up\up}+{\rm i}\ket{\up\dw\dw\up})$\\
$4$&$1$&$0$&T$_{\rm y}$&$\frac{1}{2}(\ket{\up\up\dw\dw}+
{\rm i}\ket{\dw\up\up\dw}-\ket{\dw\dw\up\up}-{\rm i}\ket{\up\dw\dw\up})$\\
$4$&$1$&$0$&T$_{\rm z}$&$\frac{1}{\sqrt{2}}(\ket{\up\dw\up\dw}-
\ket{\dw\up\dw\up})$\\
$4$&$0$&$0$&E$_1$&$\frac{1}{2}(\ket{\up\up\dw\dw}-\ket{\dw\up\up\dw}+
\ket{\dw\dw\up\up}-\ket{\up\dw\dw\up})$\\
$4$&$0$&$0$&E$_2$&$\frac{1}{\sqrt{8}}(\ket{\up\up\dw\dw}+
\ket{\dw\up\up\dw}+\ket{\dw\dw\up\up}+\ket{\up\dw\dw\up}-
2\ket{\up\dw\up\dw}-2\ket{\dw\up\dw\up})$\\ \hline
\end{tabular}
\caption{Symmetrized, non-trivial spin states of minimal
$S_z$-component for $N=3$ and $N=4$. Spin states of larger
$|S_z|$ are obtained easily.\label{spinstates}}
\end{table}\end{center}
\end{widetext}

\subsubsection{$N=3$}
In this case $\sum_{i\ne j}\varrho_{ij}^2=0$ and, in
Appendix~\ref{n3n4operators}, we replace $|\varrho_{ij}|^2$ by
$r^2$ for $N=3$ where $r$ is the mean inter-electron separation.
Then, the only non-vanishing term $H^{(2)[2,1]}$ takes the value

$\sum_{i\ne j}H_{ij}^{(2)}=\frac{r^2}{4}\sum_{i\ne j} {\rm
e}^{{\rm i}2\vartheta_{ij}}\hat S_{-i}\hat S_{-j}+{\rm h.c.}$
where $\vartheta_{ij}=0,\frac{2\pi}{3},\frac{-2\pi}{3}$ is the
azimuthal angle of $\bm{r}_{ij}$.

Thus, $H^{\rm D}$ necessarily changes $S_z$ by $\pm 2$ and has
non-vanishing matrix elements only between the A and the E
states $\{S=3/2,S_z=3/2\}\leftrightarrow\{S=1/2,S_z=-1/2\}$ and
$\{S=3/2,S_z=-3/2\}\leftrightarrow\{S=1/2,S_z=+1/2\}$ of
Table~\ref{spinstates}. Their value emerges as:
\[
\bra{\phi_{\{{\rm A},S_z=\pm 3/2\}}}H^{\rm D}
\ket{\phi_{\{{\rm E}_{\rm a,b},S_z=\mp 1/2\}}}=
-\frac{3\sqrt{3}}{4}\frac{\gamma^2}{r^3}\;.
\]
In particular, this means that the not Zeeman aligned states of
$S=3/2$ with $S_z=\pm 1/2$ remain unaffected from dipolar decay.

\subsubsection{$N=4$}
For a square arrangement of $N=4$ electrons two distances occur:
$r$ along one edge and $\sqrt{2}r$ across the diagonal.
Inspecting all the terms $H^{(0)[3,1]}\ldots H^{(2)[2,2]}$ for
$N=4$ in Appendix~\ref{n3n4operators} reveals that only contributions
$\frac{1}{16}\sum_{i\ne j} \Ke{\varrho_{+ij}^2\pm\varrho_{+\bar
i\bar j}^2} \Ke{\hat S_{-i}\hat S_{-j}\pm\hat S_{-\bar i}\hat
S_{-\bar j}}+{\rm h.c.}$ remain non-vanishing. Again, $S_z$ has
to change by $\pm 2$. The term
$\sim\Ke{\varrho_{+ij}^2+\varrho_{+\bar i\bar j}^2}$ vanishes
across the diagonal of the square, while
$\Ke{\varrho_{+ij}^2-\varrho_{+\bar i\bar j}^2}=0$ along any
edge. An evaluation yields the non-zero matrix elements between
symmetrized spin states of Table~\ref{spinstates}, i.e.\
\begin{eqnarray*}
\d\bra{\phi_{\{{\rm A},S_z=\pm 2\}}}H^{\rm D}
\ket{\phi_{\{{\rm T}_{\rm z},S_z=0\}}}&=&
-\d\frac{3}{\sqrt{2}\cdot 32}\frac{\gamma^2}{r^3}\\
\d\bra{\phi_{\{{\rm A},S_z=\pm 1\}}}H^{\rm D}
\ket{\phi_{\{{\rm T}_{\rm z},S_z=\mp 1\}}}&=&
-\d\frac{3}{64}\frac{\gamma^2}{r^3}\\
\d\bra{\phi_{\{{\rm A},S_z=\pm 2\}}}H^{\rm D}
\ket{\phi_{\{{\rm E}_1,S_z=0\}}}&=&
-\d\frac{3}{32}\frac{\gamma^2}{r^3}\;.
\end{eqnarray*}
All other matrix elements vanish identically. In particular,
states of symmetries T$_{\rm x}$, T$_{\rm y}$, E$_2$, and A with
$S_z=0$ do not exhibit dipolar decay.

\subsection{Mixed spin states}
Non-vanishing elements of $H^{\rm D}$ slightly mix eigenstates
$\ket{\phi_n}$ of $H_0$. Here we are interested in admixtures
to spin states $S$
\begin{equation}\label{admixture}
\ket{\psi_{n_S^{}}}=\ket{\phi_{n_S^{}}}+\sum_{S'\ne S}\sum_{n_{S'}^{}}
\frac{\bra{\phi_{n_{S'}^{}}}H^{\rm D}\ket{\phi_{n_{S}^{}}}}
{E_{n_{S}^{}}-E_{n_{S'}^{}}}\ket{\phi_{n_{S'}^{}}}
\end{equation}
arising from other spins $S'\ne S$. In Eq.~(\ref{admixture}) we
have disregarded the very unlikely case of accidental degeneracy
between eigenlevels of $H_0$ (cf.\ (\ref{h0})) of different
spins.\cite{degeneracy}

Eventually, this mixing will cause spin changing transitions and
thus spin relaxation. We disregard dipolar admixtures from
other states $\ket{\phi_{n_{S}'}}$ of the same spin in
Eq.~(\ref{admixture}) as those occur much more efficiently by
phonons, see in the subsequent Section. To this end, we take
$\{\ket{\psi_{n_S^{}}}\}$ as exact eigenstates of $H_0+H^{\rm
D}$. From now on we denote by $n_S^{}$ the subset of $n$-values
enumerating eigenstates of $H_0$ that belong to the definite
spin $S$.

\section{Spin relaxation rates}\label{spinrelaxrates}
Typically, the electron-phonon interaction $H^{\rm el-ph}$
establishes thermal equilibrium between electron and lattice
reservoirs on time-scales short compared to the times on which
spin changing transitions occur. This is so because the latter
cannot be achieved directly by $H^{\rm el-ph}$ (cf.\
Sect.~\ref{electronphonon}), so that equilibrium will be
established rapidly only among dot levels of given total spins.
This suggests to divide the total Hilbert space:
\[
{\cal H}=\bigoplus_S{\cal H}_S
\]
of {\em coupled\/} electron-phonon states into orthogonal
subspaces ${\cal H}_S$, labeled according to the {\em electron}
spin $S$. Transitions among subspaces occur only slowly by
the action of $H^{\rm D}$ while thermal equilibrium resides
within each of the subspaces after much shorter times $\tau^{\rm
el-ph}$ at the lattice temperature $(k_{\rm B}\beta)^{-1}$.
Consider a certain electronic spin state $S'$, as it may have
been prepared, for example, using electronic transport
techniques.\cite{elzerman05,weinprl,weinannalen} Then, the rate
\begin{equation}\label{rss}
R_{S\leftarrow S'}=\frac{\rm d}{{\rm d}t}\langle P_S(t)\rangle_{S'}
{\mbox{\Large$|$}}_{t\mbox{\tiny$\gtrsim$}\tau^{\rm el-ph}}
\end{equation}
for its decay into a particular spin $S\ne S'$ is given as
the temporal increase of the spin $S$-population $\langle
P_S(t)\rangle_{S'}$, assuming an initial (i.e.\ after
intra-${\cal H}_{S'}$ equilibration has taken place) $S'$
thermal equilibrium state,
\[
P_{S'}{\rm e}^{-\beta H}P_{S'}/
{\rm Tr}\{P_{S'}{\rm e}^{-\beta H}P_{S'}\}\;.
\]
Here, $P_S=\sum_{n_S^{}}\ket{\phi_{n_S^{}}}
\bra{\phi_{n_S^{}}}\otimes{\rm 1\!\!1}^{\rm ph}$ projects onto
${\cal H}_S$, ${\rm 1\!\!1}^{\rm ph}$ denotes a unit operator on
the phonon space. Transition rates $R_{S\leftarrow S'}$ observe
the detailed balance condition, ensuring one vanishing
(stationary) eigenvalue of the matrix $M_{SS'}=R_{S'\leftarrow
S}-\tau_{S'}^{-1}\delta_{SS'}$, which governs the rate dynamics.
In the present context we are primarily interested in the total
decay rate of the initial spin $S'$ population; i.e.,
\begin{equation}\label{taus}
\tau_{S'}^{-1}=\sum_{S\ne S'}R_{S\leftarrow S'}\;.
\end{equation}
In Eq.~(\ref{rss}) the time evolution refers to the Hamiltonian
$H=H_0+H^{\rm D}+H^{\rm ph}+H^{\rm el-ph}$ where the
electron-phonon interaction $H^{\rm el-ph}$ will be discussed
next. This approach, in principle, accounts for rapid
thermalizing spin conserving multi-phonon transitions within
subspaces ${\cal H}_S$.

\subsection{Coupling to phonons}\label{electronphonon}
Electron-phonon coupling in semiconductors has been studied
intensively in the 1950-ies and 1960-ies. For homopolar
semiconductors, such as Si or Ge, deformation potential coupling
\cite{shockely} has been established. It can be expressed as
\cite{cardona}
\begin{equation}\label{helph}
H^{\rm el-ph}=\sum_{\bms{q}}g_{\bms{q}}\;\rho(\bm{q})\;
(b_{\bms{q}}^{}+b_{-\bms{q}}^+)
\end{equation}
where we have suppressed the phonon branch index. Considerably
below room temperature, pertinent to possible quantum computing,
optical phonons don't contribute so that $b_{\bms{q}}^+$ in
Eq.~(\ref{helph}) is meant to create a longitudinal acoustical
phonon of momentum $\bm{q}$. For excitations of the electronic
system most relevant are phonon wave lengths $2\pi c_{\rm
s}/\omega_{\rm pl}=50$~nm or $2\pi c_{\rm s}/\Delta=500$~nm,
assuming \cite{leoreview} $\omega_{\rm pl}\simeq 3$~meV and $\Delta\simeq
0.3$~meV, respectively, cf.\ Sect.~\ref{pocketstates} for the
definitions of the energies $\Delta$ and $\omega_{\rm pl}$. At
this wave lengths intra-valley scattering dominates. Its strength
\begin{equation}\label{g2deform}
g_{\bms{q}}^2=\frac{E_2^2}{2\rho_{\rm M}Vc_{\rm s}}|\bm{q}|
\end{equation}
mainly is regulated by the deformation potential constant $E_2$
for longitudinal coupling which takes values of about
\cite{landoltbornstein,cardona} 10~eV in Si. Further,
$g_{\bms{q}}^2$ depends on the mass density $\rho_{\rm M}$, the
normalization volume $V$ for the phonon modes, and on the sound
velocity $c_{\rm s}$. In Eq.~(\ref{helph}) the operator
\begin{equation}\label{rhonn}
\rho(\bm{q})=N\sum_{n_S^{},n_S'}\rho_{n_S^{}n_S'}(\bm{q})
\ket{\phi_{n_S'}}\bra{\phi_{n_S^{}}}
\end{equation}
of the total electron density may excite the correlated electron
system at non-zero $\bm{q}$, though at conserved total spin $S$
(and conserved $z$-component $S_z$). It can be decomposed into
the basis $\{\ket{\phi_n}\}$, where $\ket{\phi_n}$ and
$\ket{\phi_{n'}}$ have same spin $S$, i.e.,
\begin{eqnarray}\label{rhonnq}
\d\rho_{n_S^{}n_S'}(\bm{q})=\int{\rm d}\bm{r}{\rm e}^{{\rm i}\bms{qr}}
\int&&\!\!\!\!{\rm d}\bm{r}_2\ldots{\rm d}\bm{r}_N
\braket{\bm{r},\bm{r}_2,\ldots,\bm{r}_N}{\phi_{n_S^{}}}\nonumber\\
&&\times\braket{\phi_{n_S'}}{\bm{r},\bm{r}_2,\ldots,\bm{r}_N}\;.
\end{eqnarray}
At small $q\equiv|\bm{q}|$ these coefficients are expanded,
$\rho_{nn'}(\bm{q})=\delta_{nn'}^{}+\alpha(q\ell)^{\nu}$ where,
to lowest non-vanishing order, $\nu=1$ unless the electron
charge density distribution of the quantum dot or of the ensemble
of quantum dots is parity symmetric, in which case $\nu=2$. The
quantity $\ell$ either equals the typical distance between
electrons if $n$ and $n'$ belong to the same plasmon multiplet,
or $\ell\simeq(m\omega_{\rm pl})^{-1/2}$ for $n$ and $n'$ from
different plasmon multiplets. The magnitude of $\alpha$ can be
estimated by inserting Eq.~(\ref{phicp}) into Eq.~(\ref{rhonnq})
and using, for convenience, the density distribution
$\rho_{nn'}(\bm{r})$ in real space. This reveals that $\alpha$
is proportional to the maximum overlap between unequal pocket
states, i.e.\ $\max\limits_{p,p';p\ne p'}\braket{p}{p'}$, a
quantity which, in turn, is proportional \cite{zpb} to
$\Delta/\omega_{\rm pl}$.

\subsection{Transition rates}\label{transrate}
We are now in the position to calculate the phonon mediated
transition rate Eq.~(\ref{rss}) as a result of spin mixing,
see Eq.~(\ref{admixture}). Assuming a not too strong
electron-phonon coupling, use of standard time dependent
perturbation theory with respect to $H^{\rm el-ph}$, as
explicated in Appendix~\ref{goldenrule}, yields to leading
order the rate $R_{S\leftarrow S'}$, cf.\ Eq.~(\ref{rss}),
reading
\begin{eqnarray}\label{rateresult}
\!\!\!\!R_{S\leftarrow S'}&=&\d\frac{2\pi}{Z_{S'}}\sum_{n_S^{}n_{S'}^{}}
{\rm e}^{-\beta E_{n_{S'}^{}}}
J_{n_{S'}^{}n_S^{}}(|E_{n_{S'}^{}}-E_{n_{S}^{}}|)\nonumber\\
&\times&\d[n(|E_{n_{S'}^{}}-E_{n_{S}^{}}|)+
\Theta(E_{n_{S'}^{}}-E_{n_{S}^{}})]\;,
\end{eqnarray}
where we have defined the (temperature independent) coupled
density of phonon states for transitions between spins $S$ and
$S'$, respectively, as
\begin{eqnarray}\label{jomega}
&&J_{n_{S'}^{}n_S^{}}(\omega)=\d\sum_{\bms{q}}g_{\bms{q}}^2
\;\delta(\omega-|\bm{q}|c_{\rm s})\nonumber\\
&&\times\d\Biggl|\sum_{n_S'}\rho_{n_S^{}n_S'}(\bm{q})
\frac{\bra{\phi_{n_S'}}H^{\rm D}\ket{\phi_{n_{S'}^{}}}}
{E_{n_{S'}^{}}-E_{n_S'}}\\
&&\d\mbox{~~}-\sum_{n_{S'}'}\rho_{n_{S'}'n_{S'}^{}}(\bm{q})
\frac{\bra{\phi_{n_S^{}}}H^{\rm D}\ket{\phi_{n_{S'}'}}}
{E_{n_{S'}'}-E_{n_S^{}}}\Biggr|^2\;.\nonumber
\end{eqnarray}
In Eq.~(\ref{rateresult}) $Z_S=\sum_{n_S^{}}{\rm e}^{-\beta
E_{n_S^{}}}$ denotes the partition function inside the subspace
${\cal H}_S$, $n(\omega)=({\rm e}^{\beta\omega}-1)^{-1}$ the
Bose function, and $\Theta(x)$ the Heavyside step function.
Straightforwardly, higher order terms regarding $H^{\rm el-ph}$
can also be considered for $R_{S\leftarrow S'}$, although the
corresponding explicit expressions are rather lengthy. In
Eq.~(\ref{admixture}) we have assumed that phonon states and
energy eigenvalues remain unaffected by the weak dipolar mixing.

At low temperatures, $T\ll\Delta/k_{\rm B}$ compared to the
typical distance $\Delta$ between dot levels of same or of
different total spins inside the lowest plasmon multiplet, the
Bose factor $n(|\Delta|)\ll 1$ is small and only the ground
level $n_{0S}^{}$ will be occupied within each subspace ${\cal
H}_S$. In this temperature regime thermalization into the
global ground state $n=0_{S_0}^{}$ of spin $S_0$ will take place
exclusively through the direct process by emission of a resonant
phonon of energy $\Delta$ so that the coupled density of states
$J_{n_{0S}^{}0_{S_0}}(\Delta)$ controls the relaxation rate
$R_{S_0\leftarrow S}$. Still, a summation over excited levels
$n_{S_0}'>0$ and $n_{0S}'>n_{0S}^{}$ appears in
Eq.~(\ref{jomega}), as the lowest terms $n_{S_0}'=0$ and
$n_{0S}'=n_{0S}^{}$ cancel exactly. In
Section~\ref{electronphonon} it has been estimated that
non-diagonal coefficients
$\rho_{nn'}(\bm{q})\sim(\Delta/\omega_{\rm pl})(r\Delta/c_{\rm
s})^{\nu}$ for low energy and long wave lengths transitions;
here $r$ denotes the distance between electrons and $\nu=1$ or
$\nu=2$ in the absence or presence of parity symmetry. For the
electron-phonon coupling Eq.~(\ref{g2deform}) this results in a
spin transition rate at zero temperature through the direct
process, reading
\begin{eqnarray}\label{t0rate}
&&\d R_{S_0\leftarrow S}=2\pi J_{n_{0S}^{}0_{S_0}}(\Delta)\\[3ex]
\mbox{with}&&\nonumber\\[1ex]
&&\d J_{n_{0S}^{}0_{S_0}}(\Delta)=\frac{E_2^2\gamma^4}{c_{\rm s}^7
\rho_{\rm M}\pi^2\omega_{\rm pl}^2}N^2n_{\rm a}^2\Delta^5\;,\nonumber
\end{eqnarray}
unless this transition is not suppressed entirely for cases
discussed in Sect.~\ref{stability}. In Eq.~(\ref{t0rate}) we
have assumed for simplicity that level separations
$E_{n_{S_0}'}-E_{0_{S_0}}\sim\Delta$ and
$E_{n_{S}'}-E_{n_{0S}^{}}\sim\Delta$ both are of the order
\cite{prb93} $\Delta$. Also, we have inserted the areal density
$n_{\rm a}=r^{-2}$ of electrons, focusing on the measured
quantity in two-dimensional samples.

In Si the rate Eq.~(\ref{t0rate}) appears to be very small at
zero temperature, $\sim 10^{-7}$~s$^{-1}$ for three electrons at
densities corresponding to $r_{\rm s}=1$, and considering a
quantum dot \cite{leoreview} of $\omega_{\rm pl}=3$~meV and
$\Delta=0.3$~meV. However, this number strongly varies with
parameters as seen in Eq.~(\ref{t0rate}). Parity symmetric
quantum dots (where $\nu=2$) would suppress this decay rate even
further at small transition energies due to
$J_{n0}(\Delta)\sim\Delta^7$ in this case. These values are, of
course, considerable smaller than the decay rates estimated from
spin orbit effects,\cite{khaetskii00,khaetskii01,woods04} if
present.

They are also smaller than the rates estimated from the
hyperfine interaction with nuclei of non-zero
spin.\cite{erlingsson01,erlingsson02,abalmassov04} Particularly
in Ref.~\onlinecite{erlingsson01} a hybrid mechanism is
considered which is closely related to the one presented here
in combining the electron-phonon coupling with a spin-mixing
interaction. Transitions between total spin $S=1$ and $S=0$ of
a two-electron quantum dot are investigated. The low
temperature rate has been estimated \cite{erlingsson01} to $\sim
10^{-2}$~s$^{-1}$ for similar quantum dot parameters as above,
assuming a two-dimensional dot fabricated on the basis of
heterostructures. This rate is proportional to the number
$N_{\rm n}$ of non-vanishing nuclear spins covered by the
electron wave function. In GaAs almost every nucleus has spin
$I=3/2$. It is instructive to determine from this result
\cite{erlingsson01} a critical concentration $\tilde C_{\rm n}$
of $^{29}$Si nuclei in silicon, the only ones of non-vanishing
spin $I=1/2$, beyond which the here described dipolar mechanism
should prevail over the spin decay via nuclear spins. For a
quantum dot of the same excitation energy, the electron wave
function in natural silicon covers only about $N_{\rm n}^{\rm
Si}\approx 10^3$ of the $^{29}$Si nuclei while \cite{erlingsson02}
in GaAs $N_{\rm n}^{\rm GaAs}\approx 10^5$. Two further important
differences between Si and GaAs have to be taken into account.
Firstly, the type of electron-phonon coupling which is
piezo-elastic in GaAs while we have deformation potential
coupling in Si. Accidentally, for the here
considered quantum dot parameters (and assuming again
heterostructures and now laterally parabolic confining
potential) $J_{nn'}(\omega)$ in Si is only by 0.6 smaller than
in GaAs. Secondly, the nuclear spin $I^{\rm Si}=1/2$ of
$^{29}$Si as compared to $I^{\rm GaAs}=3/2$ in GaAs which
reduces the coupling by $I^{\rm Si}(I^{\rm Si}+1)/I^{\rm
GaAs}(I^{\rm GaAs}+1)=1/5$. This yields $\tilde C_{\rm n}^{\rm
Si}\approx 2\times 10^{-4}$~nm$^{-3}$ (note that $C_{\rm n}^{\rm
Si}\approx 2.5\times 10^{-3}$~nm$^{-3}$ has been reported
\cite{tyryshkin} experimentally). This value is less stringent
than the isotopic purification required for the quantum
computer,\cite{kane98} based on the nuclear spins of $^{31}$P
donors, where $C_{\rm n}$ should be smaller than
$N^{-1}\:10^{-4}$~nm$^{-3}$ in Si with $N$ being the number of
qubits.

\subsection{Temperature Dependence}\label{temperaturedependence}
Through the marked increase of
$J_{n0}(\omega)\sim\omega^{4+2\nu}$ as a function of transition
energy $\omega$, relaxation can take advantage from spin
conserving thermal excursions to plasmonic excited levels and
accomplish the spin transition at an elevated energy. In
NMR-theory this possibility is called the `Orbach process'
\cite{slichter} and shows up in a steeply increasing relaxation
rate with temperature. Our formulation, Eq.~(\ref{rateresult}),
of the transition rate explicitly incorporates such thermal
excursions. They turn out to influence considerably the
temperature dependence of $R_{S_0\leftarrow S}$ which only at
low temperatures follows the Bose-behavior $\sim
J_{n0}(\Delta)n(\Delta)$ of direct transitions. Already at
temperatures not much exceeding $\Delta$ the severely stronger
increase $\sim J_{n0}(\omega_{\rm pl})\;\exp(-\omega_{\rm
pl}/k_{\rm B}T)$, following from Eq.~(\ref{rateresult}), can
easily enhance the transition rate by three orders of
magnitudes, depending on system parameters. A similarly
pronounced increase of (nuclear) spin relaxation rates has been
discussed in detail \cite{zpbconv} in the context of quantum
rotating molecules: substantial increases in spin changing
transition rates by more than six orders of magnitudes are
depicted with Figure~3b of Ref.~\onlinecite{zpbconv}. Again,
the stronger increasing density of coupled phonon states in
quantum dots of parity symmetry should lead to even more
pronounced temperature sensitivity.

\section{Resum\'e}\label{resume}
We have investigated dipolar interactions between the magnetic
moments of electrons confined to one or to several quantum dots
and studied the rate of inelastic total spin changing
transitions. As compared to the coupling to nuclear spins
\cite{erlingsson01,erlingsson02,khaetskii02,abalmassov04,dassarma}
and to spin-orbit induced
decay,\cite{khaetskii01,kavokin,woods04,golovach04,badescu} dipolar
spin decay turns out as much weaker. However, either of the earlier
studied mechanisms can, at least in principle, be eliminated by
a suitable sample design. It is therefore possible that the
dipolar interactions between electronic spin moments, together
with the coupling to lattice modes,\cite{sarmacomment} as
discussed in the present work, will ultimately limit long time
quantum computations, even when devices become optimally designed.
Experimentally, the dipolar mechanism should show up most
directly by observing at low temperatures the dependence on the
electron density, cf.\ Eq.~(\ref{t0rate}).

Upon generalizing previous approaches we incorporate here
electron-electron interactions \cite{kavokin,badescu} in the
quantum dot(s) which, additionally, are important for magnetic
features.\cite{reimann,epl00,mikhailov,jhjwh,mak} For example,
ground state spins may exceed the values $S_0=0$ or $S_0=1/2$
expected for non-interacting electrons. We focus on the limit
of strong interactions, where electronic many-body wave functions
can be described as `pocket states' \cite{fkp,zpb,annalen} and
where the spectrum exhibits spin-split plasmon multiplets. The
dipolar interaction is decomposed according to the symmetric
group and non-vanishing matrix elements are determined in their
dependence on spatial parts of the collective electron wave
functions. As we have shown, particular spatial symmetries of
the quantum dot(s) can reduce the number of non-zero elements.

One important result is the stability of $N=2$ electron spins.
Irrespective of the dimensionality, the shape of the quantum
dot(s), or of the electron-electron interaction strength, the
decay of triplet states into the singlet ground state is always
suppressed. Any dipolar decay channel will require
participation of further electrons. This is important, for
example, for two coupled qubits, the basic (gate)-element for
quantum information processing.

Also at larger electron numbers, non-ground state spins can be
stable with respect to dipolar interactions. We have discussed
the case of $N=3$ electrons on an equilateral triangle. Here,
$S_z=\pm 1/2$ states of the $S=3/2$ sub-manifold prove robust
against decay into any $S=1/2$ state. Only spin polarized
$S_z=\pm 3/2$ states decay into the $S=1/2$ sub-manifold.
Further, $N=4$ electrons on a square shaped quantum dot exhibit
robust $S=1$ states of T$_{\rm x}$ and T$_{\rm y}$ symmetry, and
$S=0$ states of E$_2$ symmetry.

Owing to the smallness of magnetic and in particular dipolar
energies, compared to dot level separations, the energy
accompanied with an actual spin transition has to be provided by
the reservoir of lattice vibrations. As in previous approaches
\cite{khaetskii00,khaetskii01,erlingsson01,golovach04,woods04}
we have considered the coupling to acoustic phonons. Parity
symmetric dots are weaker coupled to phonons, which further
suppresses spin decay in this case. Additionally, we have
accounted for rapid thermal excursions of the system {\em
within\/} electron-phonon subspaces of given (many electron) dot
spins. This enables one to deduce the dependence of spin
relaxation over a wider range of temperatures as compared to the
resonant direct process. As a result we found a very striking
increase of the spin decay rate. This rate grows with
temperature considerably steeper than the naively expected
proportionality to the Bose function describing direct
processes: It occurs already at temperatures that barely exceed
the energy difference between the lowest levels of different
spins, but is still considerably smaller than the energy for
plasmon excitations. Although we find amazingly stable spin
configurations at low temperatures this marked temperature
sensitivity restricts the operation temperatures of quantum
computing dots (unless quantum computation can be confined to
the stable spin configurations) to values that are not exceeding
much the lowest level separations.

Because the same phonon energy reservoir is considered in
previous work
\cite{khaetskii00,khaetskii01,erlingsson01,erlingsson02,golovach04,woods04}
for spin decay, a similar scenario regarding the dot symmetries
and the spin conserving phonon induced excursions should apply
also to magnetic mechanisms of the spin-orbit or of the hyperfine
type. We expect therefore a similarly striking temperature sensitivity
as obtained here for these mechanisms as well.

\begin{acknowledgements}
This work has been supported by the DFG Collaborative Research
Center SFB 631: ``Solid State Based Quantum Information Processing:
Physical Concepts and Materials Aspects'', project A5.
\end{acknowledgements}
\begin{widetext}
\begin{appendix}
\section{Symmetrized operators inducing spin changing transitions
for $N=3$ and $N=4$ electrons}\label{n3n4operators}
The following explicit form of dipolar operators
Eq.~(\ref{decompose}) contain non-vanishing elements for $N=3$:
\begin{displaymath}\begin{array}{rcl}
H^{(0)[2,1]}&=&\d\sum_{i\ne j}H_{ij}^{(0)}-\frac{1}{6}
\Biggl\{\Ke{\frac{1}{4}\sum_{i\ne j}|\varrho_{ij}|^2}
\Ke{\sum_{i\ne j}(\hat S_{+i}\hat S_{-j}+\hat S_{-i}\hat S_{+j})}\\[3ex]
&&\hspace*{2cm}\d\mbox{}+\Ke{\sum_{i\ne j}\zeta_{ij}^2}
\Ke{\sum_{i\ne j}\hat S_{zi}\hat S_{zj}}\Biggr\}\\[3ex]
H^{(1)[2,1]}&=&\d\sum_{i\ne j}H_{ij}^{(1)}-\frac{1}{12}
\Biggl\{\Ke{\sum_{i\ne j}\zeta_{ij}\varrho_{-ij}}
\Ke{\sum_{i\ne j}(\hat S_{+i}\hat S_{zj}+\hat S_{zi}\hat S_{+j})}\\[3ex]
&&\hspace*{2cm}\d\mbox{}+\Ke{\sum_{i\ne j}\zeta_{ij}\varrho_{+ij}}
\Ke{\sum_{i\ne j}(\hat S_{-i}\hat S_{zj}+
\hat S_{zi}\hat S_{-j})}\Biggr\}\\[3ex]
H^{(2)[2,1]}&=&\d\sum_{i\ne j}H_{ij}^{(2)}-\frac{1}{24}
\Biggl\{\Ke{\sum_{i\ne j}\varrho_{+ij}^2}
\Ke{\sum_{i\ne j}\hat S_{-i}\hat S_{-j}}+
\Ke{\sum_{i\ne j}\varrho_{-ij}^2}
\Ke{\sum_{i\ne j}\hat S_{+i}\hat S_{+j}}\Biggr\}
\end{array}\end{displaymath}
and for $N=4$:
\begin{displaymath}\begin{array}{rcl}
H^{(0)[3,1]}&=&\d\frac{1}{4}\sum_{i\ne j}\Biggl\{
\Ke{\frac{|\varrho_{ij}|^2-|\varrho_{\bar i\bar j}|^2}{4}}
\Ke{(\hat S_{+i}\hat S_{-j}+\hat S_{-i}\hat S_{+j})-
(\hat S_{+\bar i}\hat S_{-\bar j}+\hat S_{-\bar i}\hat S_{+\bar j})}\\[3ex]
&&\hspace*{1cm}\d\mbox{}+\Ke{\zeta_{ij}^2-\zeta_{\bar i\bar j}^2}
\Ke{\hat S_{zi}\hat S_{zj}-\hat S_{z\bar i}\hat S_{z\bar j}}\Biggr\}\\[3ex]
H^{(1)[3,1]}&=&\d\frac{1}{8}\sum_{i\ne j}\Biggl\{

\Ke{\zeta_{ij}\varrho_{-ij}-\zeta_{\bar i\bar j}\varrho_{-\bar i\bar j}}
\Ke{(\hat S_{+i}\hat S_{zj}+\hat S_{zi}\hat S_{+j})-
(\hat S_{+\bar i}\hat S_{z\bar j}+\hat S_{z\bar i}\hat S_{+\bar j})}\\[3ex]
&&\hspace*{1cm}\d\mbox{}+\Ke{\zeta_{ij}\varrho_{+ij}-
\zeta_{\bar i\bar j}\varrho_{+\bar i\bar j}}
\Ke{(\hat S_{-i}\hat S_{zj}+\hat S_{zi}\hat S_{-j})-
(\hat S_{-\bar i}\hat S_{z\bar j}+
\hat S_{z\bar i}\hat S_{-\bar j})}\Biggr\}\\[3ex]
H^{(2)[3,1]}&=&\d\frac{1}{16}\sum_{i\ne j}\Biggl\{
\Ke{\varrho_{+ij}^2-\varrho_{+\bar i\bar j}^2}
\Ke{\hat S_{-i}\hat S_{-j}-\hat S_{-\bar i}\hat S_{-\bar j}}
+\Ke{\varrho_{-ij}^2-\varrho_{-\bar i\bar j}^2}
\Ke{\hat S_{+i}\hat S_{+j}-\hat S_{+\bar i}\hat S_{+\bar j}}\Biggr\}\\[3ex]
H^{(0)[2,2]}&=&\d\frac{1}{4}\sum_{i\ne j}\Biggl\{
\Ke{\frac{|\varrho_{ij}|^2+|\varrho_{\bar i\bar j}|^2}{4}}
\Ke{(\hat S_{+i}\hat S_{-j}+\hat S_{-i}\hat S_{+j})+
(\hat S_{+\bar i}\hat S_{-\bar j}+\hat S_{-\bar i}\hat S_{+\bar j})}\\[3ex]
&&\hspace*{1cm}\d\mbox{}+\Ke{\zeta_{ij}^2+\zeta_{\bar i\bar j}^2}
\Ke{\hat S_{zi}\hat S_{zj}+\hat S_{z\bar i}\hat S_{z\bar j}}\Biggr\}\\[3ex]
&&\d\mbox{}-\frac{1}{12}\Biggl\{\Ke{\frac{1}{4}\sum_{i\ne j}|\varrho_{ij}|^2}
\Ke{\sum_{i\ne j}(\hat S_{+i}\hat S_{-j}+\hat S_{-i}\hat S_{+j})}+
\Ke{\sum_{i\ne j}\zeta_{ij}^2}\Ke{\sum_{i\ne j}\hat S_{zi}\hat S_{zj}}
\Biggr\}\\[3ex]
H^{(1)[2,2]}&=&\d\frac{1}{8}\sum_{i\ne j}\Biggl\{
\Ke{\zeta_{ij}\varrho_{-ij}+\zeta_{\bar i\bar j}\varrho_{-\bar i\bar j}}
\Ke{(\hat S_{+i}\hat S_{zj}+\hat S_{zi}\hat S_{+j})+

(\hat S_{+\bar i}\hat S_{z\bar j}+\hat S_{z\bar i}\hat S_{+\bar j})}\\[3ex]
&&\hspace*{1cm}\d\mbox{}+\Ke{\zeta_{ij}\varrho_{+ij}+\zeta_{\bar i\bar j}
\varrho_{+\bar i\bar j}}\Ke{(\hat S_{-i}\hat S_{zj}+\hat S_{zi}\hat S_{-j})+
(\hat S_{-\bar i}\hat S_{z\bar j}+
\hat S_{z\bar i}\hat S_{-\bar j})}\Biggr\}\\[3ex]
&&\d\mbox{}-\frac{1}{24}\Biggl\{\Ke{\sum_{i\ne j}\zeta_{ij}\varrho_{-ij}}
\Ke{\sum_{i\ne j}(\hat S_{+i}\hat S_{zj}+\hat S_{zi}\hat S_{+j})}\\[3ex]
&&\hspace*{1cm}\d\mbox{}+\Ke{\sum_{i\ne j}\zeta_{ij}\varrho_{+ij}}
\Ke{\sum_{i\ne j}(\hat S_{-i}\hat S_{zj}+
\hat S_{zi}\hat S_{-j})}\Biggr\}\\[3ex]
H^{(2)[2,2]}&=&\d\frac{1}{16}\sum_{i\ne j}\Biggl\{
\Ke{\varrho_{+ij}^2+\varrho_{+\bar i\bar j}^2}
\Ke{\hat S_{-i}\hat S_{-j}+\hat S_{-\bar i}\hat S_{-\bar j}}+
\Ke{\varrho_{-ij}^2+\varrho_{-\bar i\bar j}^2}
\Ke{\hat S_{+i}\hat S_{+j}+\hat S_{+\bar i}\hat S_{+\bar j}}\Biggr\}\\[3ex]
&&\d\mbox{}-\frac{1}{48}\Biggl\{\Ke{\sum_{i\ne j}\varrho_{+ij}^2}
\Ke{\sum_{i\ne j}\hat S_{-i}\hat S_{-j}}+\Ke{\sum_{i\ne j}\varrho_{-ij}^2}
\Ke{\sum_{i\ne j}\hat S_{+i}\hat S_{+j}}\Biggr\}\;.
\end{array}\end{displaymath}
In the above expression, $(\bar i,\bar j)$ take the two values
out of $1,\ldots,4$ that are both different from $(i,j)$.
\protect\section{Derivation of Eq.~(\ref{rateresult})}\label{goldenrule}
In a perturbative expansion w.r.t.\ $H^{\rm el-ph}$ of either of
the two time evolution operators appearing in Eq.~(\ref{rss}) we
write
\begin{eqnarray}\label{intpict}
\d{\rm e}^{-{\rm i}Ht}&=&\d{\rm e}^{-{\rm i}\tilde H_0t}
\Biggl[{\rm 1\!\!1}^{\rm ph}-{\rm i}\int_0^t{\rm d}t'\;{\rm e}^{{\rm i}\tilde
H_0t'}H^{\rm el-ph}{\rm e}^{-{\rm i}\tilde H_0t'}\\[3ex]
&-&\d\int_0^t{\rm d}t'\int_0^{t'}{\rm d}t''\;{\rm e}^{{\rm i}\tilde
H_0t'}H^{\rm el-ph}{\rm e}^{-{\rm i}\tilde H_0(t'-t'')}
H^{\rm el-ph}{\rm e}^{-{\rm i}\tilde H_0t''}+\ldots\Biggr]\;.\nonumber
\end{eqnarray}
Here, $\tilde H_0=H_0+H^{\rm D}+H^{\rm ph}$ with $H_0$ defined
in Eq.~(\ref{h0}) and the eigenstates of $H_0+H^{\rm D}$ taken
according to Eq.~(\ref{admixture}). To second (i.e.\ lowest
non-vanishing) order in $H^{\rm el-ph}$ only two of the second
terms in the square bracket of (\ref{intpict}) contribute to
Eq.~(\ref{rss}), yielding with (\ref{helph})
\begin{eqnarray*}
&&\d R_{S\leftarrow S'}=\frac{1}{Z_{S'}}\sum_{\bms{q}}g_{\bms{q}}^2
\sum_{n_S^{}n_{S'}}{\rm e}^{-\beta E_{n_{S'}}}
\frac{\rm d}{{\rm d}t}\int_0^t{\rm d}t'\int_0^t{\rm d}t''\;
{\rm i}\:\bra{\phi_{n_{S'}}}{\rm e}^{{\rm i}(H_0+H^{\rm D})t'}
\rho(\bm{q}){\rm e}^{-{\rm i}(H_0+H^{\rm D})t'}\ket{\phi_{n_S^{}}}\\[3ex]
&&\d\times(-{\rm i})\:\bra{\phi_{n_S^{}}}{\rm e}^{{\rm i}(H_0+H^{\rm D})t''}
\rho(\bm{q}){\rm e}^{-{\rm i}(H_0+H^{\rm D})t''}\ket{\phi_{n_{S'}}}
\Ke{\bar n_{\bms{q}}{\rm e}^{{\rm i}c_{\rm s}|\bms{q}|(t'-t'')}+
(\bar n_{\bms{q}}+1){\rm e}^{-{\rm i}c_{\rm s}|\bms{q}|(t'-t'')}}\;.
\end{eqnarray*}
Here, the Bose factors $\bar n_{\bms{q}}=({\rm e}^{\beta
c_{\rm s}|\bms{q}|}-1)^{-1}$ result after thermal averaging over
phonon modes. Inserting now eigenstates $\ket{\psi_n}$ of
$(H_0+H^{\rm D})$ for the $\ket{\phi_n}$, according to
(\ref{admixture}), and carrying out the long time limit
$t\to\infty$, yields for the decay rate
\begin{eqnarray*}
\d R_{S\leftarrow S'}&=&\d\frac{2\pi}{Z_{S'}}\sum_{\bms{q}}g_{\bms{q}}^2
\sum_{n_{S^{}}n_{S'}}{\rm e}^{-\beta E_{n_{S'}}}\\[3ex]
&&\times\d\Biggl[{\rm i}\Kr{\bra{\psi_{n_{S'}}}-\sum_{n_{S''}}
\frac{\bra{\phi_{n_{S'}}}H^{\rm D}\ket{\phi_{n_{S''}}}}
{E_{n_{S'}}-E_{n_{S''}}}\bra{\psi_{n_{S''}}}}\rho(\bm{q})\\[3ex]
&&\rule{3ex}{0ex}\times\d\Kr{\ket{\psi_{n_{S^{}}}}-\sum_{n_{S''}}
\frac{\bra{\phi_{n_{S''}}}H^{\rm D}\ket{\phi_{n_{S^{}}}}}
{E_{n_{S^{}}}-E_{n_{S''}}}\ket{\psi_{n_{S''}}}}\Biggr]\\[3ex]
&&\times\d\Biggl[-{\rm i}\Kr{\bra{\psi_{n_{S^{}}}}-\sum_{n_{S''}}
\frac{\bra{\phi_{n_{S^{}}}}H^{\rm D}\ket{\phi_{n_{S''}}}}
{E_{n_{S^{}}}-E_{n_{S''}}}\bra{\psi_{n_{S''}}}}\rho(\bm{q})\\[3ex]
&&\rule{3ex}{0ex}\times\d\Kr{\ket{\psi_{n_{S'}}}-\sum_{n_{S''}}
\frac{\bra{\phi_{n_{S''}}}H^{\rm D}\ket{\phi_{n_{S'}}}}
{E_{n_{S'}}-E_{n_{S''}}}\ket{\psi_{n_{S''}}}}\Biggr]\\[3ex]
&&\times\d\Ke{\bar n_{\bms{q}}
\delta(E_{n_{S'}}-E_{n_{S^{}}}+c_{\rm s}|\bm{q}|)+(\bar n_{\bms{q}}+1)
\delta(E_{n_{S'}}-E_{n_{S^{}}}-c_{\rm s}|\bm{q}|)}\;,
\end{eqnarray*}
which is readily brought into the form (\ref{rateresult}) with
(\ref{jomega}), after employing (\ref{rhonn}) and using the
orthogonality of spin states. Further calculational details can
be found in Ref.~\onlinecite{zpbconv}.
\end{appendix}
\end{widetext}


\begin{thebibliography}{10}
\bibitem{sarmacomment} Note that the dipolar interaction alone,
in the absence of phonons, would slightly mix spins but would not
cause spin decay, cf.\ Ref.~\onlinecite{dassarma}.
\bibitem{entanglement} Quantum computing employs entanglement of
different total spin states, such as the singlet and the triplet
state in the simplest situation of two electron spins, at Zeeman
energies that are comparable to the level separations between correlated
electron energies.
\bibitem{degeneracy} Exact degeneracy would entail
straightforward degenerate perturbative treatment (which poses
no conceptual difficulty when defining new `effective spins');
this case is, however, beyond the scope of the present work.
\end{thebibliography}
\end{document}